\documentclass[aps,preprint,showpacs,preprintnumbers,amsmath,amssymb,floatfix]{revtex4}
\usepackage[german,english]{babel}
\usepackage{bm,amsmath,amsfonts}
\usepackage{amssymb}
\usepackage{graphicx}

\def\beq{\begin{equation}}
\def\eeq{\end{equation}}
\def\beqn{\begin{eqnarray}}
\def\eeqn{\end{eqnarray}}

\begin{document}

\title{Optimal time-dependent lattice models for nonequilibrium dynamics}
\author{Kaspar Sakmann\footnote{E-mail: kaspar.sakmann@pci.uni-heidelberg.de}, 
Alexej I. Streltsov,
Ofir E. Alon\footnote{Present address: Department of Physics, University of Haifa at Oranim, Tivon 36006, Israel}, 
and Lorenz S. Cederbaum}
\affiliation{Theoretische Chemie, Physikalisch-Chemisches Institut, Universit\"at Heidelberg,\\
Im Neuenheimer Feld 229, D-69120 Heidelberg, Germany}
\begin{abstract}
Lattice models are central to the physics of ultracold atoms and condensed matter.
Generally, lattice models contain {\em time-independent} 
hopping and interaction parameters that are derived 
from the Wannier functions of the noninteracting problem. 
Here, we present a new concept based 
on {\em time-dependent} Wannier functions and the variational principle 
that leads to optimal time-dependent lattice models.
As an application, we use the Bose-Hubbard model
with time-dependent Wannier functions to study an interaction quench scenario
involving higher bands. We find a separation of time-scales in the 
dynamics.  The results are compared 
with numerically exact results
of the time-dependent many-body Schr\"odinger equation. 
We thereby show that -- under some circumstances -- 
the multi-band nonequilibrium dynamics of a quantum system 
can be obtained essentially at the cost of a single-band model. 
\end{abstract}

\maketitle

Quantum phenomena of particles in periodic and quasi-periodic potentials are
central themes in theoretical physics. 
Alone the question about the nature of 
the many-body ground state in such 
potentials has been investigated for decades \cite{qpts}. 
Ultracold atoms in optical lattices are highly controllable realizations of
such many-body systems and allow to directly monitor their 
nonequilibrium dynamics. Moreover, they promise
to provide insight into the physics of real solids.
Lattice models play a crucial role in the 
current physical understanding of such systems. 
At the heart of a lattice model is the idea of lattice site localized orbitals which are 
commonly referred to as Wannier functions
\cite{Wannier,SolidStatebook}.
In a lattice model Wannier functions are used as a single-particle basis for the 
many-body Hamiltonian.
Among the most famous lattice models are certainly
the single-band bosonic \cite{BHmodel} and fermionic 
\cite{HJ}
versions of the Hubbard model, both of which predict phase 
transitions between insulating and superfluid phases.
Recently there has been large interest in the multi-band physics of ultracold atoms 
by both theorists and experimentalists alike
\cite{ASC05a,ScD05,IsG05,JYD06, SWL08,LBS08, BWS09,LCM09,BJJ,WBS10}.
Interparticle interactions already couple the ground state
to higher bands \cite{ASC05a, JYD06, LBS08}, and it 
was shown recently that even very weak interactions 
lead to multi-band physics in nonequilibrium problems \cite{BJJ}. 

Multi-band phenomena seem to be abundant and we would like 
to address the question of the optimal way to deal with them theoretically.
This calls for optimal lattice models
that incorporate multi-band physics effectively. Here, we present a 
completely new concept that leads to variationally optimal lattice models for 
multi-band nonequilibrium dynamics and statics problems.
Our idea is to allow the Wannier functions of a given lattice model to depend on 
time and to determine the many-body dynamics
from the variational principle.
The concept is applicable to bosons and fermions as well as
any number of lattice sites, particles and bands.
As an application, we use the Bose-Hubbard 
model with time-dependent Wannier functions 
to study a sudden quench of the interaction in a double-well problem that involves higher bands.
We find a separation of time-scales in the dynamics.
On short time-scales higher-band excitations 
lead to rapid density oscillations, not captured on the single-band level.
On longer time-scales we observe oscillations between 
condensed and fragmented states. The results are compared 
with numerically exact results
of the time-dependent many-body Schr\"odinger equation. 
We thereby show that when the lattice is sufficiently deep and the interaction couples to higher bands the multi-band nonequilibrium dynamics of a quantum system 
can be obtained essentially at the cost of a single-band model. 
Already the physics of this example proves to be much richer 
than what the single-band Bose-Hubbard (BH) model can access.

Consider a one-dimensional (1D) lattice of $M$ sites. 
We make the following {\em ansatz} for the 
time-dependent Wannier functions 
\begin{equation}\label{tdworbitals}
w_k(x,t)=\sum_{\alpha=1}^\nu d_{k}^{\alpha}(t)w^{\alpha}_k(x),\,\,\,\,\,\,k=1,\dots,M,
\end{equation}
where $w^{\alpha}_k(x)$ denotes the conventional Wannier function 
of the band $\alpha$ at lattice site $k$
and the $d_{k}^{\alpha}(t)$ its time-dependent amplitude. 
If $\nu$ is restricted to one, the ansatz (\ref{tdworbitals}) 
reduces to the conventional  Wannier functions of the lowest band.
Using (\ref{tdworbitals}) as a basis, the ansatz for the  many-boson wave function becomes
$\vert\Psi(t)\rangle=\sum_{\vec{n}}C_{\vec{n}}(t)\left|\vec n ;t\right>$, where 
the sum is over all permanents $\left|\vec n ;t\right>=\vert n_1, n_2,\dots,n_M;t\rangle$ 
of $N$ bosons residing in $M$ time-dependent Wannier functions. 
We write $\hat b_k^\dagger(t)$ ($\hat b_k(t)$) for the operator that creates (annihilates) a boson
in the time-dependent Wannier function $w_k(x,t)$ and 
$\hat n_k(t) = \hat b_k^\dagger(t)\hat b_k(t)$ for the corresponding occupation number operator.
The bosonic commutation relations $[b_k(t),b_q^\dagger(t)]=\delta_{kq}$ hold at all times.
For simplicity we use dimensionless units in which $\hbar=m=1$.
Analogous to the derivation of the BH model, the Hamiltonian
of the time-dependent BH (TDBH) model 
\begin{eqnarray}\label{deftdbhham}
\hat H_{TDBH}&=&\sum_{k=1}^M \biggl\{-J_{kk+1}(t)\hat b_k^\dag(t) \hat b_{k+1}(t)-J_{kk+1}^\ast(t) \hat b_{k+1}^\dag(t) \hat b_k(t)\nonumber\\
 && +\epsilon_k(t)\hat n_k(t) + \frac{U_{kkkk}(t)}{2}\left[\hat n_k(t)\left(\hat n_k(t)-1\right)\right]\biggr\}
\end{eqnarray}
can then be derived from the full many-body Hamiltonian $\sum_{i=1}^N h(x_i)+\lambda_0\sum_{i<j}^N\delta(x_i-x_j)$.
Here, $h(x)=-\frac{1}{2}\frac{\partial^2}{\partial x^2} + V(x)$ is the one-body part of the Hamiltonian including 
the external potential $V(x)$.
The time-dependent 
parameters in Eq. (\ref{deftdbhham}) are defined as $J_{kk+1}(t)=-\int dx w^\ast_k(x,t)h(x) w_{k+1}(x,t)$, 
$\epsilon_k(t)=\int dx w^\ast_k(x,t )h(x)w_k(x,t)$ 
and $U_{kkkk}(t)=\lambda_0\int dx\vert w_k(x,t)\vert^4$.
Contrary to the BH model, the hopping parameter $J_{kk+1}(t)$, the one-body 
energy $\epsilon_k(t)$ and the interaction parameter $U_{kkkk}(t)$ 
are now time-dependent and can vary from one lattice site to another. 
Note also that the hopping parameter $J_{kk+1}(t)$ will generally not be real. 
It is then possible to derive coupled equations of motion for the time-dependent Wannier functions and the coefficients
of the many-body wave function from the variational principle \cite{TDVPbook1,MCTDHB2}:
\begin{eqnarray}\label{tdbh_orb_eqs}
i\vert\dot w_k(t)\rangle&=&
\hat P_k(t)\left[\hat h \vert w_k(t)\rangle + \sum_{l=k\pm 1}^M\frac{\rho_{kl}(t)}{\rho_{kk}(t)} \hat h \vert w_l(t)\rangle
+ \frac{\rho_{kkkk}(t)}{\rho_{kk}(t)} \hat U_{kk}(t)\vert w_k(t)\rangle\right],\nonumber\\
i\dot{\bf C} (t)&=&{\bf H}(t){\bf C}(t),
\end{eqnarray}
where ${\bf C}(t)=\{C_{\vec n}(t)\}$. 
The derivation and a discussion of Eqs. (\ref{tdbh_orb_eqs}) are given in the appendix.
The operators
$\hat{P}_k(t)= \sum_{\beta=1}^\nu \vert w_{k}^{\beta}\rangle\langle w_{k}^{\beta}\vert -  \vert w_{k}(t)\rangle\langle w_{k}(t)\vert$
appearing in Eq. (\ref{tdbh_orb_eqs}) are projection operators, $U_{kk}(x,t)=\lambda_0 \vert w_k(x,t)\vert^2$ and
${\bf H}_{\vec n \vec n'}(t)=\left<\vec n ;t\left|\hat H_{TDBH}\right| \vec n' ;t\right>$.
Similarly, the quantities
$\rho_{kl}(t)=\langle\Psi(t)\vert \hat b_k^\dagger(t)\hat b_l(t)\vert\Psi(t)\rangle$ and 
$\rho_{kkkk}(t)=\langle\Psi(t)\vert \hat b_k^\dagger(t)\hat b_k^\dagger(t)\hat b_k(t)\hat b_k(t)\vert\Psi(t)\rangle$ 
are matrix elements of the first- and second-order reduced density matrices. 
We stress that the optimal lattice model governed by Eqs.~(\ref{tdworbitals}-\ref{tdbh_orb_eqs})
preserves all conservation laws.
In particular, particle-number is a good quantum number,
energy is conserved when the original Hamiltonian is time-independent, 
and spatial symmetries of this Hamiltonian
are preserved in time for initial conditions preserving them.
Summarizing the above, we have 
obtained the TDBH model for which  
the wave function and the parameters of the Hamiltonian (\ref{deftdbhham})
evolve according to the variational principle.
The dynamics is determined by the coupled equations of motion (\ref{tdbh_orb_eqs}).

Now we would like to demonstrate the above ideas in an apparently simple two-site example. 
The dynamics of bosons on two lattice sites has recently been 
studied intensively using the BH model
\cite{Lee06,DHC07,K08,C08,G08,W08,B09,S09,WaG09,G09,BMC10}.
A question that arises naturally in dynamical problems is the response
of a quantum system to a perturbation. When the system is in the ground state and the perturbation
is realized as a variation of one of the parameters in the Hamiltonian, the problem is known 
as a quantum quench. Quenches have recently received growing attention
in the context of quantum gases in lattices \cite{KLA07,KLR07,RDO08,Rou10}.
Generally, the quench dynamics is studied using single-band Hubbard models
or simplifications thereof. Here, we study a quench going beyond  the BH model
and would like to investigate what new physics appears.
Here,  $V(x)=V_{0} \cos^2(x)$ is an external potential on an interval of length $2\pi$
that realizes a ring lattice of $M=2$ sites (denoted $L$ and $R$) 
with $V_{0}=25E_r$, where $E_r=1/2$ is the recoil energy. The splitting $2J=2.08 \times 10^{-3}$ of the ground  
and the first excited single-particle state of $V(x)$ determines the Rabi oscillation period $t_{Rabi}=\pi/J$. 
To characterize the interaction strength 
we use the parameter $\lambda=\lambda_0(N-1)$. 
The quench is then realized as a sudden change from 
$\lambda=0$ to $\lambda=0.6$ in a system of $N=20$ bosons, 
initially prepared in the noninteracting ground state of $V(x)$.
Within the BH model this quench corresponds to a change from $U/J=0$ to $U/J=25.8$.
We will now study and compare the dynamics
of the TDBH and the BH model. For the TDBH model the number of bands $\nu$
used for the expansion of each time-dependent Wannier function in Eq. (\ref{tdworbitals})
is increased until convergence is reached. 
In the present example
we found that the results for $J(t), U(t)$, the one-particle density and the momentum distribution do not change visibly anymore 
for $\nu\simeq5$, and $\nu=10$ was used throughout for full convergence.  
Furthermore, the spatial symmetry of the problem implies that only 
bands with even Wannier functions, $\alpha=1,3,5,7,9$, 
contribute. Moreover, $J(t)$ remains real, $\epsilon_L(t)=\epsilon_R(t)\equiv\epsilon(t)$, 
and $U_{LLLL}(t)=U_{RRRR}(t)\equiv U(t)$
at all times. All TDBH computations were carried out on a standard desktop computer 
and none of the  computations took longer than a few hours.

\begin{figure}[]
    \centering
    \includegraphics[height=11cm,angle=0]{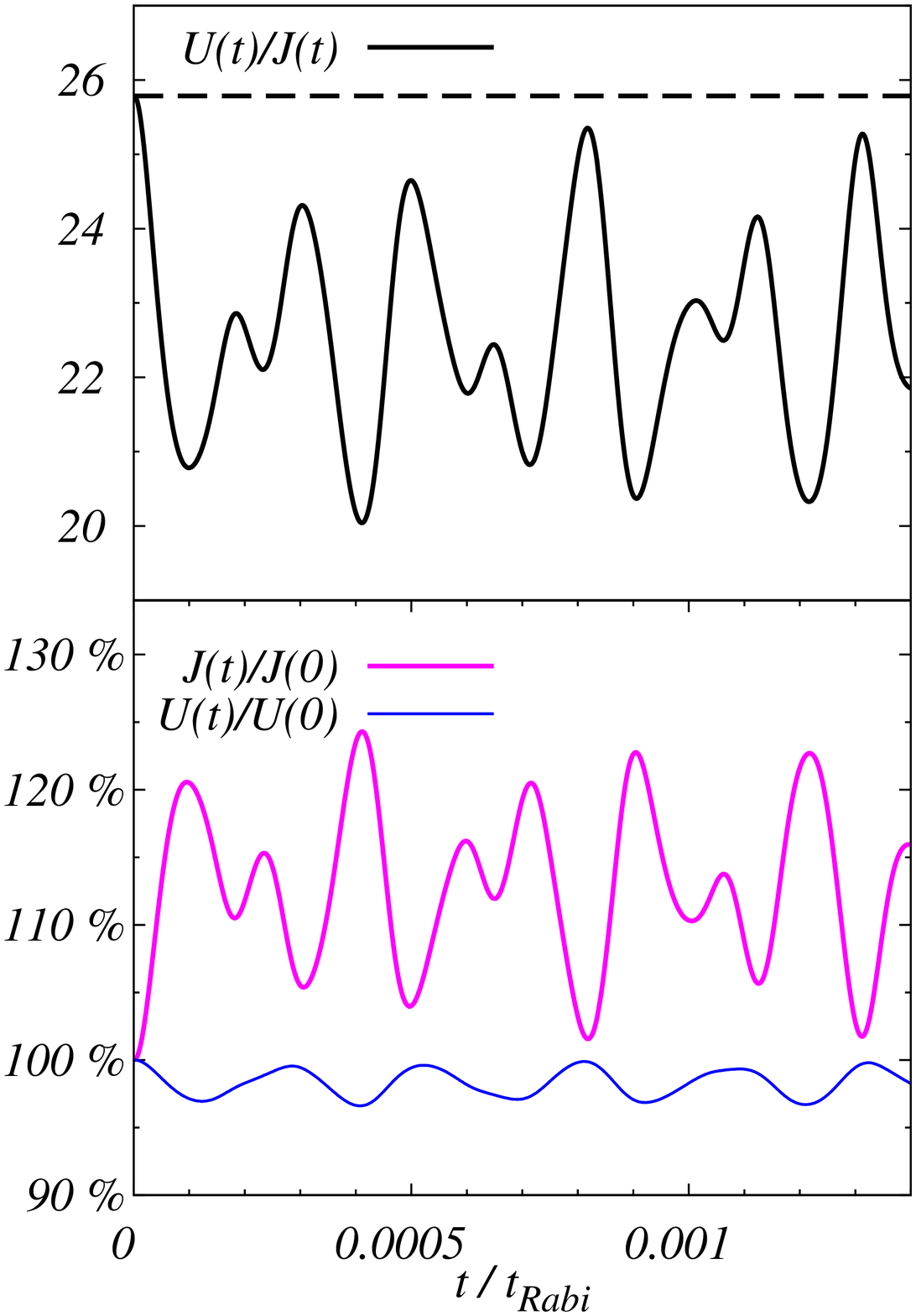}
    	\caption{(color online). Bose-Hubbard model with time-dependent Wannier functions (TDBH model). Shown is the 
		 time-dependence of the  hopping and the interaction parameters $J(t)$ and $U(t)$
	         for a quench. Initially a system of $N=20$ bosons is prepared in the noninteracting ground state 
		 of a two-site ring lattice, see text for details. At $t=0$ the interaction strength 
		 is switched on to $U(0)/J(0)=25.8$ and the resulting dynamics is monitored.
		 Top: $U(t)/J(t)$ oscillates rapidly with a large amplitude between about $25.8$ and $20$. 
		 Note the time-scale. $U(t)/J(t)$ is always smaller than its initial value 
		 which is the value of the BH model (dashed line). 
		 Bottom: $J(t)$ and $U(t)$ oscillate in time. $J(t)$ is always larger,
		 $U(t)$ always smaller than its initial value. The oscillations are due to
		 multi-band excitations induced by the quench. 
		 $J(t)$ varies over almost $25\%$, $U(t)$ over about $4\%$ 
		 of its initial value. The hopping process is obviously much more involved than what 
		 the time-independent parameter $J$ of the BH model can capture.  
	         The respective BH results are constant in time and equal to the values at $t=0$.
		 All quantities shown are dimensionless.
		  }
    \label{UandJ}
\end{figure}

On physical grounds it can be anticipated that the sudden increase of the (repulsive) 
interaction strength will lead to breathing oscillations 
of the one-particle density $\rho^{(1)}(x\vert x;t)$ in each of the wells. 
However, as the final interaction strength 
is not very strong these density oscillations should
have a small amplitude only. 
It is important to note that for the BH model the ratio of the interaction and the hopping parameter
$U/J$ is the only relevant parameter and also 
that it is constant in time after the quench.
In the TDBH model on the other hand we expect $U(t)/J(t)$ to vary, 
because the Wannier functions can change in time. This is indeed the case.
Fig. \ref{UandJ} shows the parameters
$U(t)$ and $J(t)$ of the TDBH model and their ratio $U(t)/J(t)$  as a function of time.
As expected, right after the quench the density in each well 
expands and hence $U(t)$ decreases, while $J(t)$ increases. 
Both parameters then oscillate at a high frequency.
Note the short time-scale.
The interaction broadens the density and therefore
$U(t)$ ($J(t)$) is always  smaller (larger) than its value at time $t=0$. 
As $J(t)$ depends on the overlap of adjacent time-dependent Wannier functions,
it is very sensitive to any variation of their tails. 
It varies over almost  $25\%$! The interaction parameter $U(t)$ is 
sensitive to variations of the bulk density and varies only over about $4\%$. 
As a result, the ratio $U(t)/J(t)$ of the TDBH model varies 
between $25.8$ and $20$, a range of about $25\%$.
In contrast, for the BH model the ratio $U/J=25.8$ is constant at all times as mentioned above. 
The rapid oscillations of $U(t)$ and $J(t)$ are only possible 
because the Wannier functions can evolve in the space 
of multiple bands and thus represent multi-band excitations. 
This clearly demonstrates the advantages of the TDBH model over the BH model
in reproducing the physics of the quench studied here.

\begin{figure}[]
    \centering
    \includegraphics[width=9cm,angle=-90]{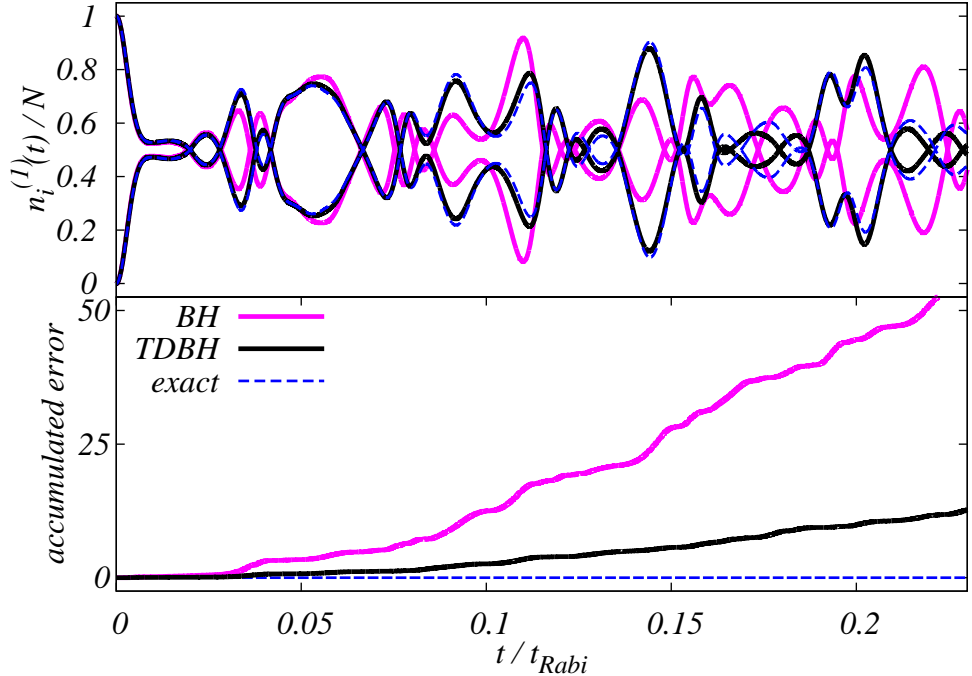}
    	\caption{(color online). Quench dynamics in a double well as in Fig. \ref{UandJ}. 
                 Top: shown are the natural occupations of the BH (solid magenta lines) model, the TDBH (solid black lines) model
	         and the exact (dashed blue lines) result as a function of time.  Starting from a fully condensed  state 
	         the dynamics is intricate and shows oscillations 
		 between partially condensed and fragmented states. The TDBH result follows the exact one
	         closely and in particular reproduces the frequencies of the oscillations well.
                 The BH result deviates quickly from the exact one.
                 Bottom: the accumulated error (see text) of the TDBH model is always smaller than that of the BH model
	         and grows much more slowly. All quantities shown are dimensionless.
}
    \label{tdbh-quench-0-to-06-potential25Er-natoccsM4}
\end{figure}

We now turn to the question about the nature of the quantum state after the quench
and to longer time-scales. We first focus on the eigenvalues of the first-order reduced density matrix, 
the natural occupation numbers $n_i^{(1)}(t)$, which determine the degree 
to which the system is condensed or fragmented \cite{PeO56,NoS82}.
In Fig.~\ref{tdbh-quench-0-to-06-potential25Er-natoccsM4} (top) the natural
occupation numbers of the TDBH model are 
shown together with those of the BH model. For comparison also the numerically exact dynamics 
of the time-dependent many-body Schr\"odinger equation is shown, 
obtained using the multiconfigurational time-dependent Hartree for bosons method.
For details see the literature \cite{MCTDHB1,MCTDHB2}. Note that the time-scale is now more 
than a hundred times larger compared to Fig. \ref{UandJ}.
It is clearly seen that all three results display oscillations between fragmented and partially 
condensed states. For the first few oscillations the 
three results essentially coincide, which implies that for this 
problem the natural occupation numbers are not very sensitive
to the previously discussed multi-band excitations.
However, after a few oscillations between condensed and fragmented states 
the BH result deviates substantially from the exact one.
On the other hand, the TDBH result follows the exact one closely for many oscillations. 
In  Fig.~\ref{tdbh-quench-0-to-06-potential25Er-natoccsM4}
(bottom) the accumulated error of the BH model, defined as the accumulated difference 
$\frac{1}{N}\int_0^t dt'\vert n_{1,BH}^{(1)}(t')-n_{1,exact}^{(1)}(t')\vert$ 
between the largest natural occupation number of the BH model
and the numerically exact result are shown, 
together with the  analogous quantity of the TDBH model.
The accumulated error of the TDBH model is always smaller 
than that of the BH model and grows more slowly, 
which means that also on longer time-scales
the TDBH model captures the true physics much better than the BH model.

\begin{figure}[]
    \centering
    \includegraphics[width=9cm,angle=-90]{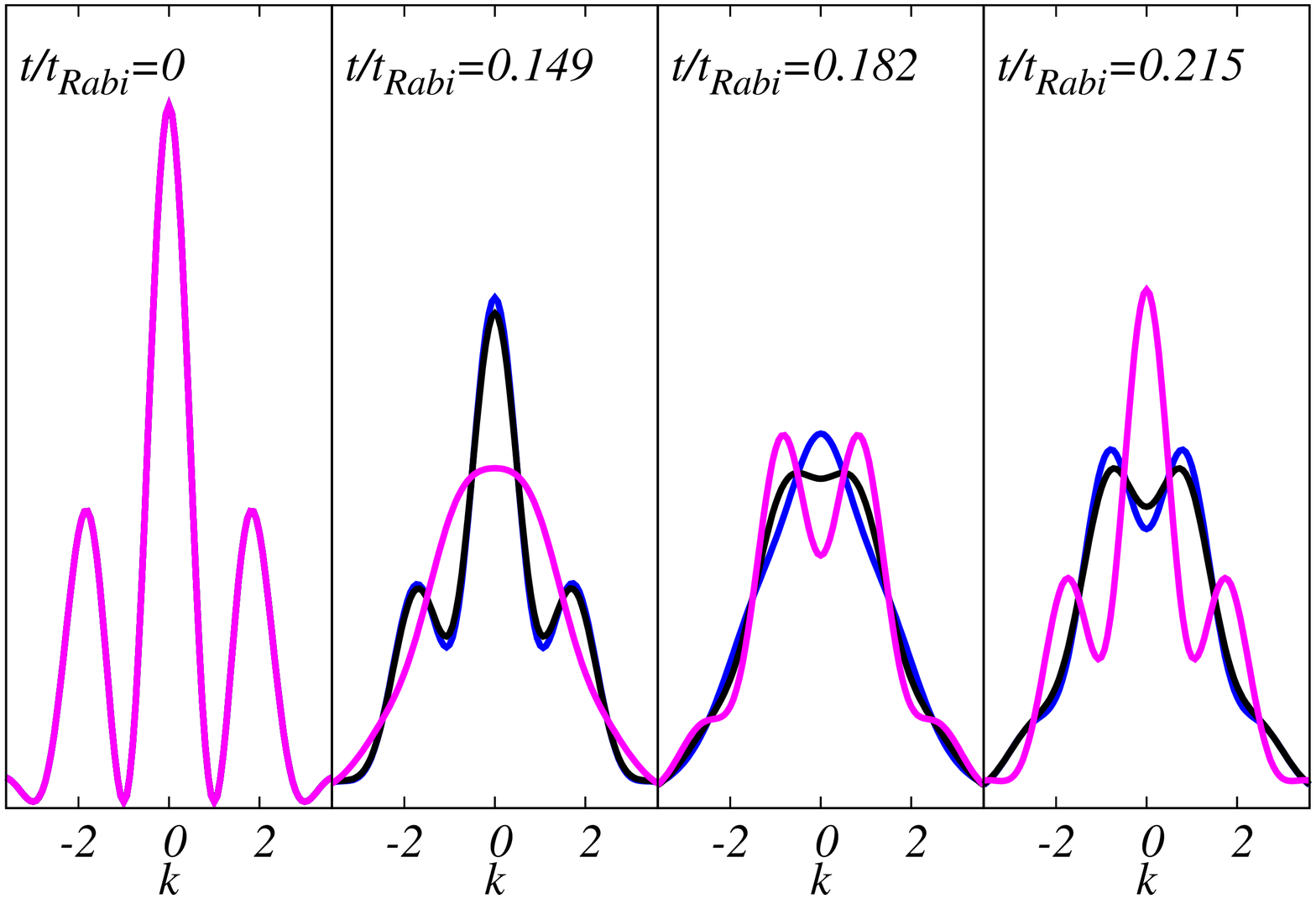}
        \caption{(color online). Quench dynamics in a double well as in Fig.~\ref{UandJ}. 
	Shown is the one-particle momentum distribution of the BH (solid magenta lines) model, 
        the TDBH (solid black lines) and the exact result (solid blue lines)
	at different times. The initial state is fully coherent and for a short time
	the three results have similar momentum distributions (not shown).
        The TDBH result is always much closer
	to the exact one than the BH result. Differences of the TDBH (BH) result relative to the exact one 
	always occur when the natural occupations differ, see Fig. \ref{tdbh-quench-0-to-06-potential25Er-natoccsM4}.
        All quantities shown are dimensionless.
 }
    \label{momentaseries}
\end{figure}

We will now discuss the implications of the differences of the natural occupation numbers
on observables. The rapid density oscillations discussed earlier 
have a very small amplitude, and consequently 
hardly any dynamics is visible in real space. 
However, a substantial dynamics occurs in momentum space.
Fig.~\ref{momentaseries} shows the one-particle 
momentum distribution $\rho^{(1)}(k\vert k;t)$ of the exact, the TDBH and the BH result at $t=0$ 
where they all coincide, and at later times where differences have developed.
The momentum distribution of the TDBH model is always
closer to the exact one and follows it for a long time, 
whereas that of the BH model deviates quickly.
Here, the TDBH model allows to obtain the precise multi-band nonequilibrium dynamics
essentially at the cost of a single-band model.
Of course, in order to obtain the exact result the terms that were neglected so far in Eq. (\ref{deftdbhham})
have to be included into the model, e.g., the terms responsible for correlated tunneling.
Thus, the advantages of the TDBH model over the BH model have been clearly proven.

In this work we have generalized the concept of lattice models 
by letting their Wannier functions become time-dependent. 
For such lattice models equations of motion can be derived from 
the variational principle, leading to variationally optimal lattice models 
that can efficiently incorporate multi-band physics.
The concept is applicable to both nonequilibrium dynamics as well as statics.
In the latter case the ground state can be computed using imaginary time-propagation.
The variational principle ensures that for identical initial conditions the optimal
lattice model improves on the original lattice model,
if the latter contains the terms 
of the full many-body Hamiltonian 
relevant to the problem at hand.
Since the numerical effort of
the optimal and standard lattice
models are comparable,
the former can also be used 
to assess the quality of the latter: if the respective results differ noticeably,
the standard lattice model is inapplicable.
As an application, we have presented the equations of motion for the
TDBH model, i.e., the Bose-Hubbard model with time-dependent Wannier functions.
We then studied a quantum quench problem in a double-well using the TDBH model. 
Multi-band excitations were found
that resulted in high-frequency, large amplitude oscillations of the
hopping and interaction parameters. Such phenomena are 
not accessible to the BH model. On a longer time-scale 
the multi-band excitations also affect the nature 
of the quantum state and the momentum distribution.
We have compared the results of the TDBH and the BH model for the natural occupation numbers and the 
momentum distributions with numerically exact ones of the many-body 
Schr\"odinger equation. For this problem, we find that
the TDBH results follow the exact ones for a long time, 
while the BH results deviate quickly.
Interestingly, the numerical effort for solving the equations of 
motion of the TDBH model is essentially the same as for the BH model, see the appendix for details. 

Summarizing, standard lattice models can be greatly extended at little extra cost 
through the use of time-dependent Wannier functions. 
It will be interesting 
to examine what other physical multi-band phenomena are 
accessible to such time-dependent lattice models, e.g., in disordered media 
or in the case of the fermionic Hubbard model.
As a further direction we suggest to combine 1D optimal lattice models  
with the recently developed adaptive 
time-dependent density-matrix renormalization group methods \cite{t1,t2,t3}
to treat non-equilibrium dynamics of larger lattice systems,
as is presently done utilizing 
standard lattice models.

\appendix*
\section{Derivation and discussion of the equations of motion of the TDBH model}\label{app}

We will now derive Eqs. (\ref{tdbh_orb_eqs}) 
and discuss their properties. Using the ansatz for the many-body wave function 
$\vert\Psi(t)\rangle=\sum_{\vec{n}}C_{\vec{n}}(t)\left|\vec n ;t\right>$ given above, 
the action functional of the TDBH model reads 
\beq\label{acfun}
 S\left[\{C_{\vec{n}}\},\{d_{k}^\alpha\}\right]=
  \int dt \biggl\{ \left<\Psi\left|\hat H_{TDBH} - i\frac{\partial}{\partial t} \right|\Psi\right>
 - \sum_{i} \mu_{i}(t)\left[\left<w_i(t)|w_i(t)\right> - 1 \right]\biggr\}, 
\eeq
where the Lagrange multipliers are introduced to ensure orthonormality 
between the time-dependent Wannier functions. Note that Eq. (\ref{tdworbitals}) implies that
time-dependent Wannier functions on different lattice sites are orthogonal by construction.
In order to take functional derivatives it is useful to express the expectation value 
$\langle\Psi\vert\hat H_{TDBH} - i\frac{\partial}{\partial t} \vert\Psi\rangle$ as an explicit function
of the amplitudes $\{d_k^\alpha(t)\}$ and expansion coefficients $\{C_{\vec{n}}(t)\}$. On the one hand 
$\langle\Psi\vert\hat H_{TDBH} - i\frac{\partial}{\partial t} \vert\Psi\rangle$ can be written as:
\begin{equation}\label{coefeqn}
\biggl<\Psi\biggl|\hat H_{TDBH} - i\frac{\partial}{\partial t} \biggr|\Psi\biggr>
=\sum_{\vec n}C^\ast_{\vec{n}}(t)\biggl[\sum_{\vec n'}\biggl<\vec n ;t\biggl| 
\hat H_{TDBH} - i\frac{\partial}{\partial t}\biggr| \vec n' ;t\biggr>
C_{\vec{n'}}(t)-i\frac{\partial C_{\vec{n}}(t)}{\partial t}\biggr].
\end{equation}
Using Eq. (\ref{coefeqn}) we now require stationarity of the action functional 
with respect to variations of the coefficients,  $0=\frac{\delta S}{\delta C^\ast_{\vec{n}}(t)}$,
and obtain:
\begin{equation}\label{tdbh_coef_eq}
\sum_{\vec n'}\left<\vec n ;t\left| 
\hat H_{TDBH} - i\frac{\partial}{\partial t}\right| \vec n' ;t\right>
C_{\vec{n'}}(t)=i\frac{\partial C_{\vec{n}}(t)}{\partial t}.
\end{equation}
On the other hand $\langle\Psi\vert\hat H_{TDBH} - i\frac{\partial}{\partial t} \vert\Psi\rangle$
can also be written as:
\begin{eqnarray}\label{orbeqn}
\left<\Psi\left|\hat H_{TDBH} - i\frac{\partial}{\partial t} \right|\Psi\right>
=
\sum_{k,q=k,k\pm 1}\sum_{\alpha,\beta}^\nu \rho_{kq}(t)\, {d_k^\alpha}^\ast(t) \left(h_{k_\alpha q_\beta}-i\delta_{kq}\delta_{\alpha\beta}\frac{\partial}{\partial t}\right) d_q^\beta(t)\nonumber \\
+\frac{1}{2}\sum_{k}\sum_{\alpha\beta\gamma\delta}\rho_{kkkk}(t)\, {d_k^\alpha}^\ast(t) {d_k^\beta}^\ast(t)\,  
U_{k^\alpha k^\beta k^\gamma k^\delta}\,d_k^\gamma(t) d_k^\delta(t) 
- i \sum_{\vec n}C_{\vec n}(t)\frac{\partial C_{\vec n}^\ast(t)}{\partial t}, 
\end{eqnarray}
where $h_{k^\alpha q^\beta}=\langle w_k^\alpha\vert\hat h\vert w_q^\beta\rangle$, 
$U_{k^\alpha k^\beta k^\gamma k^\delta}=\lambda_0\int dx {w_k^\alpha}(x)^\ast {w_k^\beta}(x)^\ast w_k^\delta(x)w_k^\delta(x)$
and we have used $\langle w_k^\alpha\vert w_q^\beta\rangle =\delta_{kq}\delta_{\alpha\beta}$.
Note that the matrix elements of the first- and second-order reduced density matrices are 
functions of the coefficients $\{C_{\vec n}(t)\}$ only.
Using Eq. (\ref{orbeqn})  we now require stationarity of the action functional with 
respect to variations of the time-dependent amplitudes, 
$0=\frac{\delta S}{\delta {d^\alpha_k}^\ast(t)}$, and obtain
\begin{equation}\label{orbvaria}
\sum_{q=k,k\pm1}\rho_{kq}(t)\langle w_k^\alpha\vert \hat h \vert w_q(t)\rangle
+ \rho_{kkkk}(t) \langle w_k^\alpha\vert\hat U_{kk}(t)\vert w_k(t)\rangle =\mu_{k}(t) \langle w_k^\alpha\vert w_k(t)\rangle +i\rho_{kk}(t)\langle w_k^\alpha\vert\dot w_k(t)\rangle,
\end{equation}
for $\alpha=1,\dots,\nu$ and $k=1,\dots,M$, where we have used  $d_k^\alpha(t)=\langle w_k^\alpha\vert w_k(t)\rangle$. The Lagrange multipliers $\mu_{k}(t)$ can be determined from Eqs.~(\ref{orbvaria}), which upon resubstituting the result, 
multiplying each equation by $\vert w_k^\alpha\rangle$ and summing over $\alpha$ gives 
\begin{eqnarray}\label{tdbhgeneqs}
i\rho_{kk}(t)\hat P_k(t)\vert\dot w_k(t)\rangle&=&
\hat P_k(t)\left[\sum_{q=k,k\pm1}\rho_{kq}(t) \hat h \vert w_q(t)\rangle 
+ \rho_{kkkk}(t) \hat U_{kk}(t)\vert w_k(t)\rangle\right].
\end{eqnarray}
By including the additional phase factor $e^{-\int^t dt'\langle w_k(t)\vert\dot w_k(t)\rangle}$ 
into the definition of $\vert w_k(t) \rangle$, we find 
that $\langle w_k(t)\vert \dot w_k(t)\rangle=0$, and that
Eqs. (\ref{tdbhgeneqs}) and (\ref{tdbh_coef_eq}) reduce to Eqs. (\ref{tdbh_orb_eqs}). 

Let us briefly discuss the numerical advantage of using time-dependent 
Wannier functions over conventional ones. The fundamental difficulty in 
solving quantum lattice models lies in the fact 
that the dimension of the many-particle Hilbert space grows very quickly
as a function of the number of particles $N$ and the number of lattice sites $M$. 
In the case of the BH model the length of the vector ${\bf C}$
is $\binom{N+M-1}{N}$. This is also the length of 
the vector ${\bf C}$ in the  TDBH model since
we have only replaced conventional Wannier functions by time-dependent ones. However,
since each of the time-dependent Wannier functions can 
take on any shape allowed by Eq. (\ref{tdworbitals}), a much larger Hilbert 
space is available to the variational principle.
Note that the number of bands $\nu$ in which each time-dependent Wannier 
function is expanded does not enter the above dimension formula. 
If $\nu$ is restricted to one, the time-dependent  Wannier functions become time-independent
and only the second of Eqs. (\ref{tdbh_orb_eqs}) needs to be solved. In this case
Eqs. (\ref{tdbh_orb_eqs}) reduce to the equations of motion of the BH model.
If $\nu>1$ the coupled system, Eqs. (\ref{tdbh_orb_eqs}), needs to be solved. 
Then, the additional numerical cost consists of propagating not only the coefficients 
$\{C_{\vec n}(t)\}$, but also the time-dependent Wannier functions $\{w_k(x,t)\}$, i.e., $M$ 
additional nonlinear equations have to be solved.
For all but the smallest numbers of particles and lattice sites, this extra effort 
is much smaller than the effort needed to propagate the second 
of  Eqs. (\ref{tdbh_orb_eqs}). Note that the alternative of enlarging 
the Hilbert space by using multiple, say $\kappa>1$, 
bands of conventional (static) Wannier functions, 
would increase the length of the vector ${\bf C}$ 
to $\binom{N+\kappa M-1}{N}$. This illustrates why it is crucial to 
use time-dependent Wannier functions: 
time-dependent Wannier functions allow to keep the numerical effort 
of a time-dependent lattice model very close to that of the respective 
standard lattice model while efficiently incorporating multi-band physics.
Since all parameters, i.e., the coefficients $\{C_{\vec n}(t)\}$ {\em and} the Wannier functions $\{w_k(x,t)\}$ 
are at all times determined by the variational principle, 
the lattice model is optimal.

\begin{acknowledgments}
Financial support by the DFG is gratefully acknowledged.
\end{acknowledgments}

\newpage
\thispagestyle{empty}

\begin{thebibliography}{150}
\ifnum\language=1 \def\biband{und}\else\def\biband{and}\fi
\ifnum\language=1 \def\bibcomma{\ }\else\def\bibcomma{,\ }\fi

\bibitem{qpts} S. Sachdev, {\em Quantum Phase Transitions} (Cambridge University Press, 1999).

\bibitem{Wannier} G. H. Wannier, Phys. Rev. {\bf 52}, 191 (1937).

\bibitem{SolidStatebook} W. Jones and N. H. March, {\em Theoretical Solid State Physics Volume 1: Perfect Lattices in Equilibrium} (Dover, 1985).

\bibitem{BHmodel} M. P. A. Fisher {\em et al.},  Phys. Rev. B {\bf 40}, 546 (1989).  

\bibitem{HJ} J. Hubbard, Proc. R. Soc. Lond. A {\bf 276}, 238 (1963).

\bibitem{ASC05a} O. E. Alon {\em et al.}, Phys. Rev. Lett. {\bf 95}, 030405 (2005).

\bibitem{ScD05} V. W. Scarola and S. Das Sarma, Phys. Rev. Lett. {\bf 95}, 033003 (2005).

\bibitem{IsG05} A. Isacsson and S. M. Girvin, Phys. Rev. A {\bf 72}, 053604 (2005).

\bibitem{JYD06} J. Li {\em et al.}, New J. Phys. {\bf 8} 154, (2006).

\bibitem{SWL08} V. M. Stojanovi\'{c} {\em et al.}, Phys. Rev. Lett. {\bf 101}, 125301 (2008).


\bibitem{LBS08} D.-S. L\"uhmann {\em et al.}, Phys. Rev. Lett. {\bf 101}, 050402 (2008). 

\bibitem{BWS09} T. Best {\em et al.}, Phys. Rev. Lett. {\bf 102}, 030408 (2009).

\bibitem{LCM09} J. Larson {\em et al.}, Phys. Rev. A {\bf 79}, 033603 (2009).

\bibitem{BJJ} K. Sakmann {\em et al.}, Phys. Rev. Lett. {\bf 103}, 220601 (2009); 
Sakmann K, {\em et al.}, Phys. Rev. A {\bf 82} 013620 (2010).

\bibitem{WBS10} S. Will {\em et al.}, Nature {\bf 465},  197 (2010).

\bibitem{TDVPbook1} P. Kramer and M. Saraceno, {\em Geometry of the Time-Dependent Variational Principle in Quantum Mechanics} (Springer, Berlin, 1981) {\em Lecture Notes in Physics} vol {\bf 140}.

\bibitem{MCTDHB2}   O. E. Alon {\em et al.}, Phys. Rev. A {\bf 77}, 033613 (2008).



\bibitem{Lee06} C. Lee, Phys. Rev. Lett. {\bf 97}, 150402 (2006).

\bibitem{DHC07} D. R. Dounas-Frazer {\em et al.},
Phys. Rev. Lett. {\bf 99}, 200402 (2007).

\bibitem{K08} Y. Khodorkovsky {\em et al.},
Phys. Rev. Lett. {\bf 100}, 220403 (2008).

\bibitem{C08} P. Cheinet {\em et al.},
Phys. Rev. Lett. {\bf 101}, 090404 (2008).

\bibitem{G08} E. M. Graefe {\em et al.},
Phys. Rev. Lett. {\bf 101}, 150408 (2008).

\bibitem{W08} D. Witthaut {\em et al.},
Phys. Rev. Lett. {\bf 101}, 200402 (2008).

\bibitem{B09} E. Boukobza {\em et al.},
Phys. Rev. Lett. {\bf 102}, 180403 (2009).

\bibitem{S09} K. Smith-Mannschott {\em et al.},
Phys. Rev. Lett. {\bf 102}, 230401 (2009).

\bibitem{WaG09} J. Wang and J. Gong,
Phys. Rev. Lett. {\bf 102}, 244102 (2009).

\bibitem{G09} J. Gong {\em et al.},
Phys. Rev. Lett. {\bf 103}, 133002 (2009).

\bibitem{BMC10} E. Boukobza {\em et al.}, 
Phys. Rev. Lett. {\bf 104}, 240402 (2010).




\bibitem{KLA07} C. Kollath {\em et al.}, Phys. Rev. Lett. {\bf 98}, 180601 (2007).

\bibitem{KLR07} I. Klich {\em et al.},  Phys. Rev. Lett. {\bf 99}, 205303 (2007).

\bibitem{RDO08} M. Rigol {\em et al.}, Nature {\bf 452}, 854 (2008).

\bibitem{Rou10} G. Roux, Phys. Rev. A {\bf 81}, 053604 (2010).

\bibitem{PeO56} O. Penrose and L. Onsager, Phys. Rev. {\bf 104}, 576 (1956).

\bibitem{NoS82} P. Nozi\`eres, D. Saint James, J. Phys. France {\bf 43}, 1133 (1982);
		P. Nozi\`eres, in {\em Bose-Einstein Condensation}, 
		edited by A. Griffin, D. W. Snoke, and S. Stringari (Cambridge University Press, 1996) ISBN 9780521589901.

\bibitem{MCTDHB1} A. I. Streltsov {\em et al.}, Phys. Rev. Lett. {\bf 99}, 030402 (2007).

\bibitem{t1} G. Vidal, Phys. Rev. Lett. {\bf 93}, 040502 (2004).

\bibitem{t2} A. Daley {\it et al.}, J. Stat. Mech.: Theory Exp. P04005 (2004).

\bibitem{t3} S. R. White and A. E. Feiguin, Phys. Rev. Lett. {\bf 93}, 076401 (2004).

\end{thebibliography}
\end{document}